\documentclass[aps, prl, reprint, showpacs, twocolumn, 10pt, groupedaddress,noeprint]{revtex4-2}

\usepackage{graphicx, amsmath, dsfont, dcolumn, bm, times, xcolor, braket, amssymb,natbib}

\begin{document}

\title{Quantized transconductance emerges from non-symmetric quantum fluctuations: theoretical prediction}

\author{K. Mertiri, Yuli V. Nazarov}
\affiliation{Kavli Institute of Nanoscience, Delft University of Technology, 2628 CJ Delft, The Netherlands}

\date{\today}

\begin{abstract}
We show theoretically that weak quantum fluctuations induced by a non-symmetric electromagnetic environment may lead to a quantized transconductance of a multi-terminal quantum contact rather than to a blockade of transport in the contact. The result suggests the possibility to realize Quantum Hall phenomenology without its common ingredients and/or a topological quantum state.
\end{abstract}

\maketitle

The electron-electron interaction may tremendously affect electron transport in condensed matter systems. In the quantum transport paradigm, the relevant interaction affecting transport in a contact is represented by quantum fluctuations of the external electromagnetic environment \cite{QuantumTransport}.  For a tunnel contact, the effect of interaction was fully comprehended \cite{Tunnel1,Tunnel2,Tunnel3} shortly after the formulation of dissipative quantum mechanics  \cite{Legget}. It was understood that the interaction results in the suppression of tunneling, the strength of the interaction is determined by a typical dimensionless impedance $z \equiv (G_Q/2) Z $, $G_Q \equiv e^2/\pi \hbar$. In the limit $z \to \infty$ one encounters the Coulomb blockade: tunneling is fully suppressed in a finite energy interval. Remarkably, tunneling is fully suppressed at any finite $z$, even at $z \ll 1$ in the limit of low energies, that is, at vanishing temperature and voltage \cite{Tunnel1,Tunnel2,Tunnel3}, leading to an isolation of the leads. This is a case of the Anderson orthogonality catastrophe \cite{Anderson}. Later, this result was confirmed for arbitrary contact transparencies \cite{NazarovKindermann,Flensberg}.    
 
Transconductance appears in multi-terminal electric circuits. The quantization of transconductance in units of $G_Q$ has been discovered in the context of the Integer Quantum Hall Effect (IQHE) \cite{IQHE}. While generally explained as a consequence of topology, the quantization can be easily understood in the quantum transport paradigm: it corresponds to integer number of transport channels that go from one terminal to another without reflection \cite{Buttiker}. There are alternative proposals to realize quantized transconductance (QTC), such as topological multi-terminal Josephson circuits \cite{Houzet}, or quantum synchronization of Bloch and Josephson oscillations \cite{Hriscu}.

The compact way to understand the emerging isolation at $z \ll 1$ is to use the "poor man's" renormalization \cite{Legget}. At each step of renormalization, one incorporates the effect of fluctuations in the energy strip $dE$ shifting the cut-off energy accordingly. The procedure stops when the running cut-off reaches sufficiently low energies $E$ that are of the order of the voltage bias or temperature. This results in a simple differential equation for a single-channel transmission coefficient $T$ in the tunneling case $T \ll 1$, $\xi \equiv \ln(E_{\rm cut}/E)$:   
\begin{align}
\frac{ d T}{d \xi} = - 2 z T; \; T \simeq T_{\rm cut} \left(\frac{E}{E_{\rm cut}}\right)^{2z}.
\end{align}
For an arbitrary transmission coefficient, the equation reads \cite{NazarovKindermann}:
\begin{align}
\frac{ d T}{d \xi} = - 2 z T (1-T),
\end{align} 
that is, $T \to 0$ at $E \to 0$, and the only stable fixed point of the renormalization corresponds to complete isolation. Any two-terminal quantum contact can be regarded as a collection of independent channels, so the conclusion seems universal. What about multi-terminal contacts?

In this Letter, we investigate the effect of quantum fluctuations on a multi-terminal quantum contact at $z \ll 1$ in the framework of the  renormalization technique. It is essential that the impedance is now a matrix in the space of terminals $\hat Z(\omega)$. In the absence of time-reversibility, this matrix is non-symmetric in terminal indices. We reveal that isolation is not the only stable fixed point of the renormalization flow. Alternative fixed points appear at sufficiently big asymmetric part of $\hat Z$. They realize a quantized transconductance. Therefore, non-symmetric quantum fluctuations may drive a quantum contact to a state where trans-conductance is like in IQHE, this state being distinct from any IQHE state and not requiring a bulk topological insulator. 
  
\begin{figure}
    \begin{center}
    \includegraphics[width=\linewidth]{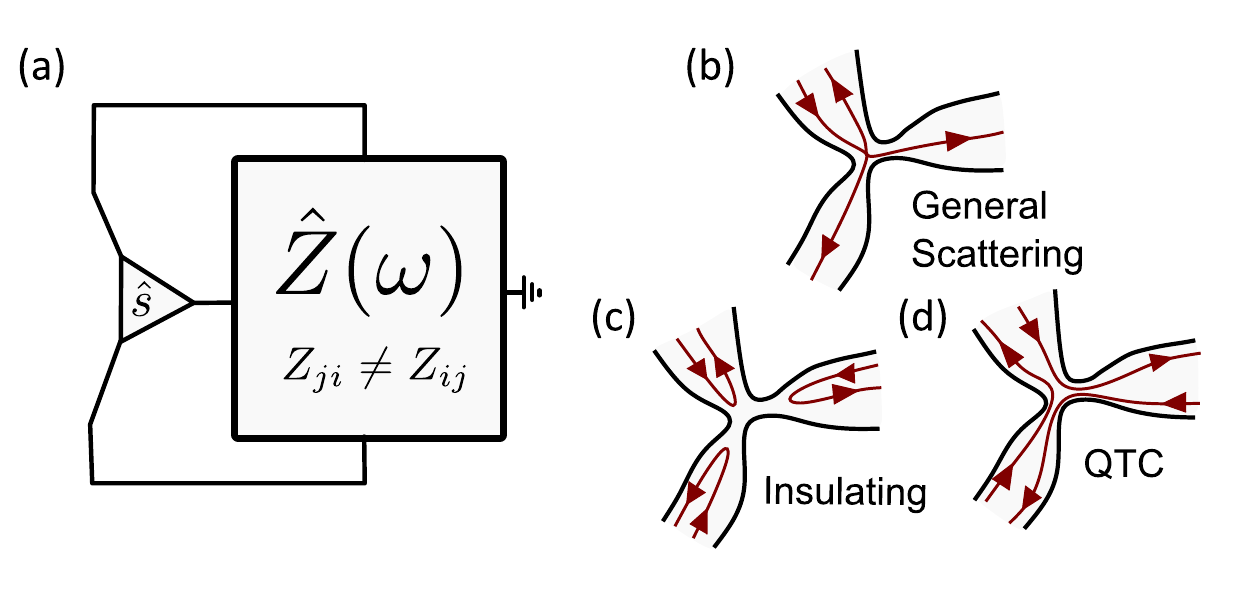}
	\end{center}
    \caption{(a) The N-terminal scatterer (N=3 in the Figure) embedded in a linear electromagnetic environment which is characterized by the generally non-symmetric $N\times N$ impedance matrix $\hat{Z}(\omega)$. (b) Sketch of a general (high-energy) scattering matrix: an incoming electron wave is scattered back and to other terminals. (c) Insulating fixed point: all incoming waves are reflected in the low-energy limit. (d) Predicted QTC fixed point: all incoming waves are fully transmitted to different terminals realizing IQHE phenomenology in the low-energy limit.}
    \label{fig: Setup}
\end{figure}

We consider the setup shown in Fig.~\ref{fig: Setup}(a). An arbitrary N-terminal contact is fully characterized by an elastic scattering matrix $\hat{s}$ (Fig.~\ref{fig: Setup}(b)). The contact is embedded in a linear electromagnetic environment characterized by the $N\times N$ impedance matrix $Z_{ij}(\omega)$. We assume that the matrix is non-symmetric. This implies the absence of time reversibility in the environment, such as that introduced by a magnetic field in the (non-quantum) Hall effect. Voltage fluctuations in the environment induce  time-dependent phases $\Phi_i$ on electron states within  terminal `$i$',  $d\Phi_i(t)/dt = V_i(t)$, and thereby cause interaction between electrons traversing the contact. We assume an equilibrium environment so that fluctuations are determined by $\hat{Z}(\omega)$.

We evaluate the Full Counting Statistics \cite{FCS} of charge transfers in the contact. As shown in Ref. \cite{NazarovKindermann}, this is a convenient way to assess the renormalization of the scattering matrix. We employ the version of  the non-equilibrium Keldysh technique \cite{Snyman} that suits the problem in hand. We concentrate on the limit of small fluctuations, $z \ll 1$, and obtain the correction to the generating function to first order in $z$. The resulting terms describe inelastic scattering processes as well as the correction to  elastic scattering amplitudes, and we concentrate on the latter. The correction involves only the {\it real} part of the impedance $\hat z = (G_Q/2) {\rm Re} \hat{Z}(\omega \to 0)$. If this matrix is frequency-independent in a wide frequency range, the correction diverges logarithmically  indicating the break-down of perturbation theory and the need of renormalization, similar to that implemented in \cite{NazarovKindermann}, but extended to the multi-terminal case. (The multi-terminal renormalization studied in Ref. \cite{LL} in Luttinger liquid setups leads to non-trivial fixed points of a different origin.)

To describe the {\it big} change of scattering, we employ the resulting renormalization flow equation for the energy-dependent scattering matrix. It involves a dimensionless variable $\xi = \ln(E_{\rm cut}/E)$, $E_{\rm cut}$ being the upper cutoff energy determined by characteristics of either the contact or the electromagnetic environment. We present the detailed derivation in the Supplemental Material \cite{supplementary}. The renormalization flow of the scattering matrix element $s_{ij}$, $i,j$ labeling channels, is given by 
\begin{equation}\label{eq: renormalization}
    \frac{ds_{ij}}{d\xi} = z_{ji}s_{ij} - \sum_{k,l} z_{kl}s_{ik} (\hat s^\dag)_{kl} s_{lj}.
\end{equation}
Here we assume for generality that there is an individual terminal for each channel, so that the matrix $\hat{z}$ is in the space of channels.

Eq.~\eqref{eq: renormalization} defines the energy dependence of the scattering matrix. Let us concentrate on the fixed points of the renormalization flow: scattering matrices in the limit $E \to 0, \xi \to \infty$ that define transport in the contact in the limit of low bias voltages and temperatures. The fixed-point $\hat{s}^{(0)}$ satisfies the following equation:
\begin{equation}\label{eq: fixedpoints}
\sum_{k} (z_{jk} - z_{ik})s_{ki}^{(0)*} s^{(0)}_{kj} = 0.
\end{equation}
We note that the insulating fixed point $\hat s^{(0)}_{ij} = \delta_{ij} e^{i \theta_i}$ (Fig.~\ref{fig: Setup}(c)) is always a valid one satisfying the above equation, $\theta_i$ being arbitrary reflection phases. The stability criterion for this point reads
\begin{equation}
\label{eq:positivity}
z_{ii} + z_{jj} - z_{ij} - z_{ji} >0 \quad \forall\, i,j. 
\end{equation}
We recognize here the general condition any physical impedance matrix should satisfy: it guarantees that the energy dissipation in the linear circuit is positive. We conclude that the insulating fixed point is {\it always} stable.  

The main message of this Letter is that there are alternative fixed points (Fig.~\ref{fig: Setup}(d)). Let us define a permutation matrix $\hat{P}$ with exactly one $P_{ij} =1$ element in each row and each column, all other elements being 0. For any $\hat{P}$,
the scattering matrix with elements $\hat s^{(0)}_{ij} = e^{i \theta'_i} P_{ij} e^{i\theta_j}$ is a fixed point, $\theta'_i,\theta_j$ being arbitrary phases in outgoing/incoming channels. It is worth mentioning that these permutation points remain fixed even if the renormalization equation contains terms of higher order in $z$ , since for these particular scattering matrices the fluctuations only affect the phases of scattering matrix elements and do not change the transmission coefficients. However, higher-order terms may affect the stability exponents and the stability thresholds that we address below.

The stability criterion is imposed on the matrix $\hat{z}\hat{P}$,
\begin{equation}
\label{eq:stability}
    (\hat{z}\hat{P})_{ii} + (\hat{z}\hat{P})_{jj} - (\hat{z}\hat{P})_{ij} - (\hat{z}\hat{P})_{ji} >0 \quad \forall\, i,j. 
\end{equation}
This criterion is challenging to fulfill (see \cite{supplementary}). For instance, all permutations containing a disjoint cycle of length 2 (full transparency in both directions) are always unstable. However, if $\hat z$ is not symmetric \cite{footnote}, some permutations may generate non-trivial stable fixed points. The scattering matrix of this permutation form implies that an incoming wave in each channel $i$ is fully transmitted to the channel $j$ defined by the permutation. This guarantees quantized transconductances.

The understanding of the renormalization flow and the stability of fixed points is facilitated by the fact that Eq.~\eqref{eq: renormalization} can be presented as
\begin{equation}
    \frac{d\hat s}{d\xi} = - \frac{\delta \varepsilon}{\delta \hat s^\dag}; \quad 
    \varepsilon(\hat s) = - \sum_{i,j} z_{ji} |s_{ij}|^2.
\end{equation}
The function $\varepsilon$ plays the role of a pseudo-potential in the space of $\hat s$. The $\xi$ derivative of $\hat s$ is proportional to the gradient of $\varepsilon$. This guarantees that $\varepsilon$ always decreases in the course of the renormalization flow. Let us exemplify the use of $\varepsilon$ for $N=3$. For the insulating fixed point, $\varepsilon_{\rm ins} = - (z_{11}+z_{22}+z_{33})$. Let us take permutation $P_1: 123 \to 312$. The corresponding $\varepsilon_{\rm P_1} = -(z_{13} + z_{21} + z_{32})$. If $\hat{z}$ is symmetric, condition \eqref{eq:positivity} implies 
$\varepsilon_{\rm P} > \varepsilon_{\rm ins}$. However, if one increases the anti-symmetric part of the matrix (which does not affect condition \eqref{eq:positivity}), one reaches $\varepsilon_{\rm P_1} < \varepsilon_{\rm ins}$. This implies the stability of this fixed point. 

There can thus be multiple stable fixed points, in which case the actual scattering matrix 
at $ E \to 0$ depends on the ``initial condition": $\hat{s}_{\rm{in}} \equiv \hat{s}(E=E_{\rm cut}) = \hat{s}(\xi=0)$.

A more common situation is that of multi-channel junctions, where there are several channels in each terminal $\alpha$. The impedance matrix acquires a block structure with $z_{ij} = z_{\alpha \beta}$ for all $i \in \alpha, j \in \beta$. 
This block structure increases the degeneracy of QTC fixed points: any matrix of the form $s^{(0)} = \hat{U} \hat{P} \hat{V}$, $\hat U,\hat V$ being arbitrary {\it block-diagonal} unitary matrices. These fixed points still correspond to a quantized transconductance. For each terminal with M channels, $m$ channels are reflected and $M-m$ are fully transmitted to other terminals. The stability of a point requires the same condition \eqref{eq:stability}.\\

From now on, let us concentrate on the minimal setup with $N=3$. The stability considerations are easy to obtain analytically for one channel in each terminal. There are only two potentially stable QTC fixed points corresponding to the  permutations $P_{1} (123 \rightarrow 312)$ and $P_{-1} \equiv P^{-1}_1$. The stability conditions involve a single parameter $Z_{a} = Z^A_{21} +Z^A_{32}+Z^A_{13}$, $\hat Z^A$ being the anti-symmetric part of $\hat Z$, and read
\begin{equation}\label{eq: stabilityconditions}
\begin{aligned}
	P_{\pm 1} \text{ is stable provided}\ \rightarrow     \pm Z_a > X_1,X_2,X_3.   
\end{aligned}
\end{equation}
Here $X_1 = Z_{11} + Z^S_{23} - Z^S_{13} - Z^S_{21}$, $\hat Z^S$ is the symmetric part of $\hat{Z}$, and $X_2,X_3$ are obtained by index permutation. Thus for any given $\hat{Z}$ either $P_{1}$ or $P_{-1}$ is stable, or both are unstable. It is possible to show that for $N=3$ the same conditions apply for any multi-channel situation as well. The only potentially stable fixed points belong to the subset $P_n$: if $n>0$, $n$ channels go from $1$ to $3$, from $2$ to $1$, from $3$ to $2$ while all other channels are reflected; if $n<0$, $-n$ channels go from $3$ to $1$, from $1$ to $2$, from $2$ to $3$. The stability thresholds are the same for all positive or negative $n$ and are given by  \eqref{eq: stabilityconditions}. The pseudo-potential reads: $\varepsilon_n = \varepsilon_{\rm ins} + (G_Q/2)(-Z_a + Z_{11}+Z_{22}+Z_{33}) n$, so all points with negative/positive $n$ have a pseudo-potential smaller than that of the insulating point upon crossing the same threshold. \\

\begin{figure}
    \centering
    \includegraphics[width=\linewidth]{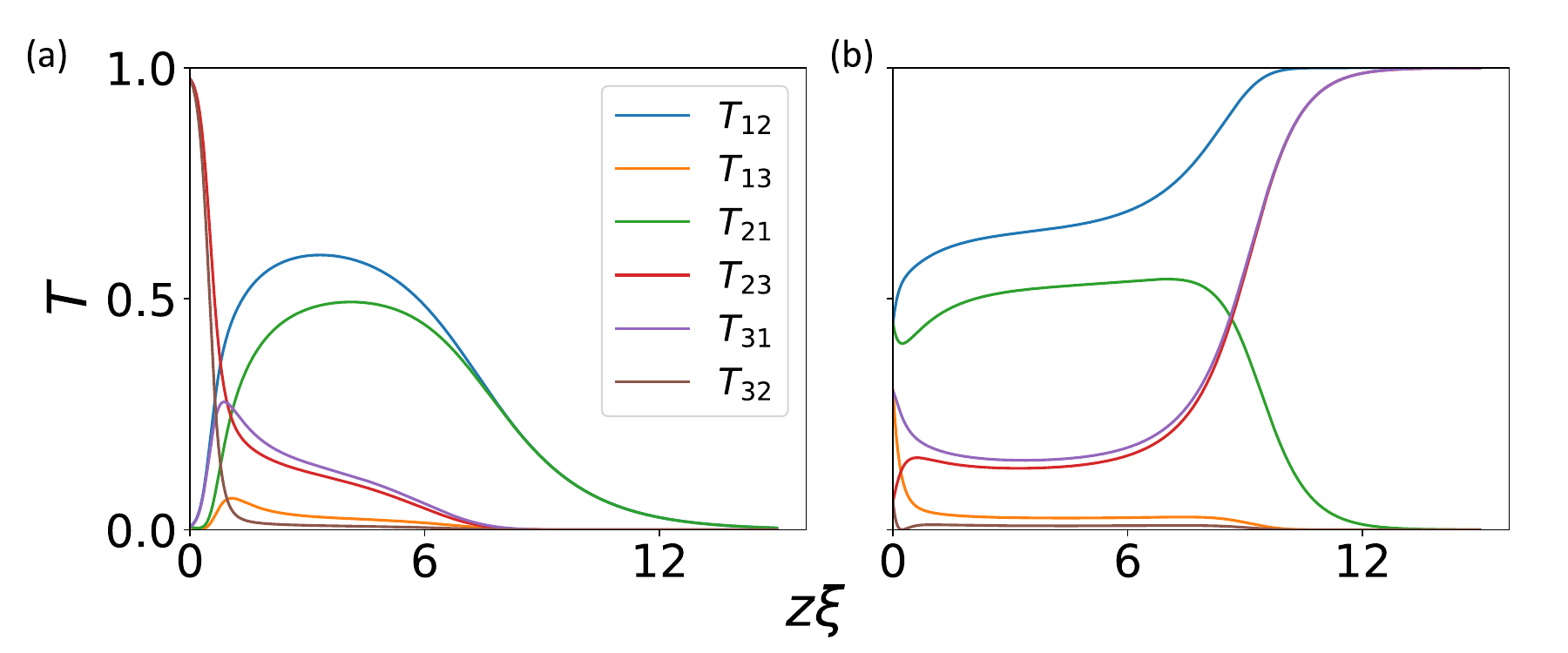}
    \caption{One channel per terminal. Energy dependence of transmission coefficients for two different choices of time-reversible $\hat{s}_{\rm in}$. (a) Realization of the insulating fixed point. (b) Realization of a QTC fixed point. We  choose $c = 1.5$ where both fixed points occur with comparable probabilities.}
    \label{fig: evolution}
\end{figure}

To illustrate and quantify the situation we perform numerical simulations for $N=3$. We fix $\hat{z}$. We pick up an ``initial" random scattering matrix $\hat{s}_{\rm in}$ from the circular unitary ensemble and evolve it with Eq.~\eqref{eq: renormalization} until a stationary $\hat{s}^{(0)}$ is achieved. (The flow induced by a non-symmetric environment does not preserve the symmetry class: an initially symmetric $\hat{s}_{\rm in}$ immediately becomes asymmetric upon the flow.) We identify the resulting fixed point and repeat the procedure accumulating the statistics of these points. Since all possible $\hat{z}$ satisfying \eqref{eq:positivity} can be realized with circuit design, for all simulations presented here we use an arbitrary symmetric, $\hat{S}$, and anti-symmetric, $\hat{A}$, $3\times 3$ matrix to define  $\hat z = z(\hat S + c\hat A$), that depends on the parameter $c$ characterizing the asymmetry. Concrete matrices in use are:
\begin{equation}
    \hat S= \begin{pmatrix}
    0.4 & 0.4 & -0.0\\
    0.4 & 0.8 & -0.6\\
    -0.0 & -0.6 & 1.6
    \end{pmatrix}, \quad
    \hat A = \begin{pmatrix}
    0 & -1.0 & 0.4\\
    1.0 & 0 & -0.8\\
    -0.4 & 0.8 & 0
    \end{pmatrix}.
\end{equation}
For this choice, $P_{\pm 1}$ is a stable point for $\pm c > c_s$, $c_s = 1.19$, while its pseudo-potential is smaller than that of the insulating point for $\pm c>c_c = 1.33$. The same applies to $P_n$.

We show in Fig.~\ref{fig: evolution} typical traces characterizing the flow of $\hat{s}(\xi)$ for the situation of multiple stable points. A choice of ``initial" $\hat{s}$ results in insulation at $E\to 0$, while another one manifests a QTC fixed point. We note a complex behavior of the transmission coefficients $T_{ij}$ in the course of the flow: they are non-monotonous, and intervals of relatively slow evolution alternate with intervals of relatively fast change. In any case, the traces saturate at the corresponding fixed points at $\xi$ several times bigger than $1/z$. 

\begin{figure}
    \centering
    \includegraphics[width=\linewidth]{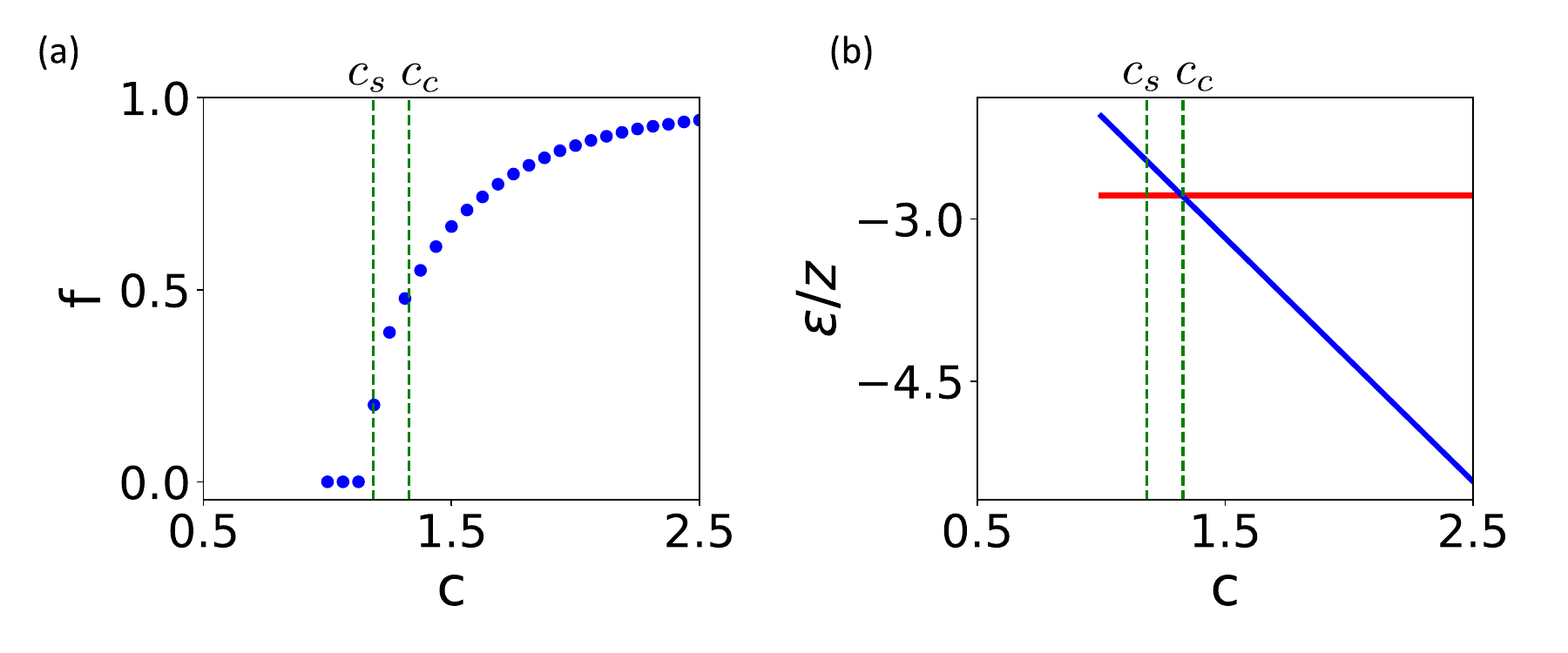}
    \caption{One channel per terminal. (a) The fraction of $\hat{s}_{\rm in}$ flowing to QTC fixed point $P_1$ versus asymmetry parameter $c$. $10^4$ runs per point in $c$. (b) The pseudo-potential $\varepsilon$ of the insulating fixed point (red) and $P_1$ (blue). $c_{s}, c_{c}$ indicate the stability and pseudo-potential crossing thresholds, respectively.}
    \label{fig: 3x1}
\end{figure}

We have accumulated the simulation results for different $c$ to assess the probabilities for the realization of various fixed points. In Fig.~\ref{fig: 3x1} we present the results for one channel per terminal. The fraction of $\hat{s}_{\rm in}$ flowing to the QTC fixed point $P_{1}$ is shown in Fig.~\ref{fig: 3x1}(a). This fraction is zero below the stability threshold $c_s = 1.19$ and increases to $1$ upon increasing $c$. At $c=c_c$, the two fixed points occur with approximately equal probabilities. To identify a point, we use the pseudo-potential achieved by the end of the renormalization flow. We plot the pseudo-potential for the two fixed points versus $c$ in Fig.~\ref{fig: 3x1}(b) to show $c_c$.

We investigate the multi-channel situation as well. For the setup with 3 channels in each of the 3 terminals, we identify 3 QTC fixed points $P_n$, $n=1,2,3$, as expected. They differ by the value of the quantized transconductances, $G_{12}=G_{23}=G_{31}=n G_Q $. Their pseudo-potentials are given in Fig.~\ref{fig: 3x3,3x10}(c). For the $\hat{z}$ of choice, all 3 points are stable at $c>c_s = 1.19$ and have pseudo-potentials lower than the insulating fixed point at $c>c_c=1.33$. In Figs.~\ref{fig: 3x3,3x10}(a),(b) we present the probabilities to achieve each of the 4 points at various $c$. The probability of the insulating point drops from $1$ to numerical zero at the threshold $c=c_s$. Right after the threshold, the probability $f_1$ of $P_1$ dominates. At $c \approx c_c$, the probabilities $f_1$ and $f_2$ are almost the same, while $f_3$ is still negligible. The latter increases and dominates upon increasing $c$, while $f_1$ and $f_2$ quench around $c \simeq 2.0$ and $c \simeq 5$, respectively. 

The numerical investigation of a big number of channels per terminal is difficult owing to the high dimension of the scattering matrix. Nevertheless, we are able to investigate the statistics of a setup with 10 channels per terminal. We expect to find 10 possible QTC fixed points with $G_{12}=G_{23}=G_{31}=n G_Q$, $n=1..10$. However, with the discretization of $c$ as in Fig.~\ref{fig: 3x3,3x10}(d) we find only the points with $n \ge 4$. Their probabilities versus $c$ display a peculiar pattern shown in Fig.~\ref{fig: 3x3,3x10}(d): each $f_n$ peaks at a certain $c_n$ and dominates in an interval around $c_n$ while quickly dropping to 0 beyond the interval. To emphasize this, we plot in the figure the fits to the data for each $f_n$ with the expression $f_n(c) = A \exp(-B(c/c_n + c_n/c - 2))$, $A,B,c_n$ being the fit parameters. 
These results motivate the following hypothesis concerning the renormalization of transconductance for a semiclassical $\hat s$ encompassing a big number of channels and correspondingly big $G \gg G_Q$. We surmise that in this case, transconductance is a quasi-smooth function of $c$, $G(c) \simeq G$. Since at each $c$ in the plot the probability is contributed by either one or two peaks, we estimate the variance $\text{Var } G(c) \simeq G_Q^2$. This looks analogous to the universal conductance fluctuations in quantum transport  \cite{QuantumTransport},  where commonly $\text{Var }  G \simeq G_Q^2$ at $\langle G \rangle \gg G_Q$. It would be interesting to research this semiclassical limit, quantifying $G(c)$ and its variation.

\begin{figure}[h]
    \centering
    \includegraphics[width=\linewidth]{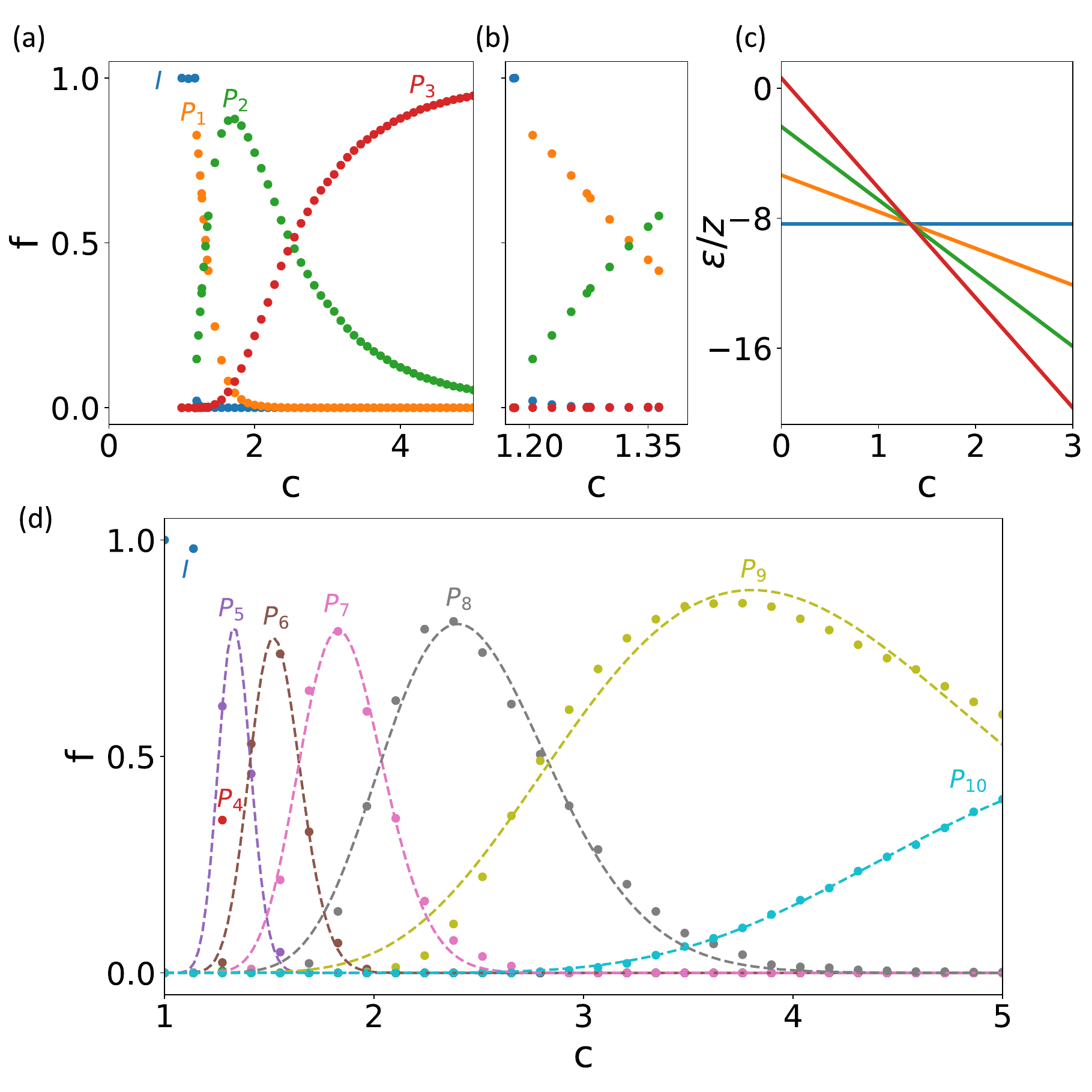}
    \caption{3 channels per terminal. (a),(b) The probabilities for the realization of QTC fixed points $P_n$ versus $c$. $10^4$ runs per point in $c$. (b) Zoom in of the same data as in (a) in the vicinity of $c_c$. (c) The pseudo-potential for $P_n$ and the insulating point. 10 channels per terminal. (d) The probabilities for the realization of QTC fixed points $P_n$ versus $c$. $10^3$ runs per point in $c$.}
    \label{fig: 3x3,3x10}
\end{figure}

{\it Experimental outlook}. It might seem that the design and realization of the linear Ohmic environment under consideration can be delegated to electronic engineers. Nevertheless, there are some points that require a prefatory discussion. 
To start with, an interesting environment can not be made by combining resistors: this would result in a circuit obeying the no-amplification property \cite{noAmplification}: $z_{ii} + z_{jk} \geq z_{ik} + z_{ji}, \forall i,j,k.$ One can show that the no-amplification property is not consistent with any alternative fixed points. The environment to be realized should be (a) amplifying, (b) asymmetric, (c) passive, (d) broadband, and (e) permitting to measure the quantum contact at low frequencies. Criteria (a) and (b) are required for alternative fixed points, while (c) and (d) are required by the renormalization approach in use. To summarize, we can satisfy all criteria using resistive elements, (on-chip) transformers \cite{networkbook} (for (a), (c), (d)), low-frequency capacitance filters (for (e)) and non-ideal Hall effect gyrators \cite{grutzmann2005hall,arlt1960halleffekt} (for (b)). The concrete circuit design and details are given in \cite{supplementary}.

{\it Conclusions and discussion}. We have demonstrated the QTC as a possible low-energy transport scenario in quantum contacts subject to non-symmetric electromagnetic fluctuations. This leads to IQHE phenomenology without 2D semiconductor heterostructures and topological states. In particular, it can be achieved in much shorter contacts. The resulting scattering matrix in combination with magnetic and superconducting leads may result in the formation of topological states and new devices associated with this. The flow equation presented is valid at $z\ll 1$, and it is feasible to complement it with higher-order corrections $\propto z^2, z^3$, etc. We note, however, that these corrections do not affect the stability of QTC points as far as they remain small.  More complex scenarios where the stability is affected at higher $Z \simeq G_Q^{-1}$ are beyond the present scope. More research is required to elucidate the effects of small non-Gaussian and non-equilibrium noise: albeit one generally expects that small perturbations do not affect the stability unless it is marginal.  An intriguing property of the renormalization is the dependence of the zero-energy regime on the high-energy scattering matrix: the latter can be changed by gate voltages resulting in sharp transitions between QTC and insulation, or between different quantized transconductance values. Such transitions can also be achieved by changing the asymmetry of $\hat{z}$. See \cite{supplementary} for a detailed illustration.

\begin{acknowledgments} 
We acknowledge useful discussions with E. Samuelsen and an important remark of D. G. Polyakov.
This research is supported by Delft Univeristy of Technology.
The data that support the findings of this article are openly available at \cite{code}.
\end{acknowledgments}

\bibliography{paper-bib}

\end{document}